%% file: KDD_Camera_Ready.tex
\documentclass[sigconf]{acmart}
\input{math_commands.tex}

\usepackage[capitalise]{cleveref}
\usepackage{soul}

\def\algname{\sc SWaT}

\AtBeginDocument{%
  }

\copyrightyear{2025}
\acmYear{2025}
\setcopyright{rightsretained}

\acmConference[KDD '25]{Proceedings of the 31st ACM SIGKDD Conference on Knowledge Discovery and Data Mining V.1}{August 3--7, 2025}{Toronto, ON, Canada} \acmBooktitle{Proceedings of the 31st ACM SIGKDD Conference on Knowledge Discovery and Data Mining V.1 (KDD '25), August 3--7, 2025, Toronto, ON, Canada} \acmDOI{10.1145/3690624.3709415} 
\acmISBN{979-8-4007-1245-6/25/08}

\makeatletter
\gdef\@copyrightpermission{
  \begin{minipage}{0.2\columnwidth}
   \href{https://creativecommons.org/licenses/by/4.0/}{\includegraphics[width=0.90\textwidth]{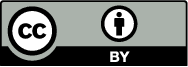}}
  \end{minipage}\hfill
  \begin{minipage}{0.8\columnwidth}
   \href{https://creativecommons.org/licenses/by/4.0/}{This work is licensed under a Creative Commons Attribution International 4.0 License.}
  \end{minipage}
  \vspace{5pt}
}
\makeatother

\begin{document}



\title{{\color{burgundy} \algname}: {{\sc{\textcolor{burgundy}{S}}}}tatistical Modeling of Video \textcolor{burgundy}{\sc{{Wa}}}tch {\textcolor{burgundy}{\sc T}}ime  through  {U}ser {Be}havior Analysis}

\author{Shentao Yang}
\authornote{Work done while authors were interns at Meta AI.}
\email{shentao.yang@mccombs.utexas.edu}
\orcid{0009-0009-8058-3149}
\affiliation{%
  \institution{University of Texas at Austin}
  \city{Austin}
  \state{TX}
  \country{USA}
}

\author{Haichuan Yang}
\email{hyang.ur@gmail.com}
\orcid{0000-0002-5714-4174}
\affiliation{%
  \institution{Meta AI}
  \city{Menlo Park}
  \state{CA}
  \country{USA}}

\author{Linna Du}
\email{linnadu@meta.com}
\orcid{0009-0009-2640-7531}
\affiliation{%
  \institution{Meta AI}
  \city{New York}
  \state{NY}
  \country{USA}
}

\author{Adithya Ganesh}
\email{acganesh@meta.com}
\orcid{0009-0001-5868-499X}
\affiliation{%
 \institution{Meta AI}
  \city{Menlo Park}
  \state{CA}
  \country{USA}
  }

\author{Bo Peng}
\authornotemark[1]
\email{peng.707@buckeyemail.osu.edu}
\orcid{0009-0000-7569-1828}
\affiliation{%
  \institution{Ohio State University}
  \city{Columbus}
  \state{OH}
  \country{USA}}

\author{Boying Liu}
\email{boyingliu@meta.com}
\orcid{0009-0005-3019-5763}
\affiliation{%
  \institution{Meta AI}
  \city{Seattle}
  \state{WA}
  \country{USA}}

\author{Serena Li}
\email{serenali@meta.com}
\orcid{0009-0008-9644-7988}
\affiliation{%
  \institution{Meta AI}
  \city{Menlo Park}
  \state{CA}
  \country{USA}}

\author{Ji Liu}
\email{ji.liu.uwisc@gmail.com}
\orcid{0000-0003-4921-4643}
\affiliation{%
  \institution{Meta AI}
  \city{Seattle}
  \state{WA}
  \country{USA}}

\renewcommand{\shortauthors}{Shentao Yang et al.}

\begin{abstract}
The significance of estimating video watch time has been highlighted by the rising importance of (short) video recommendation, which has become a core product of mainstream social media platforms.
Modeling video watch time, however, has been challenged by the complexity of user-video interaction, such as different user behavior modes in watching the recommended videos and varying watching probability over the video progress bar. 
Despite the importance and challenges, existing literature on modeling video watch time mostly focuses on relatively black-box mechanical enhancement of the classical regression/classification losses, without factoring in user behavior in a principled manner.
In this paper, we for the first time take on a user-centric perspective to model video watch time, from which we propose a white-box statistical framework that directly translates various  user behavior assumptions in watching (short) videos into statistical watch time models.
These behavior assumptions are portrayed by our domain knowledge on users' behavior modes in video watching.
We further employ bucketization to cope with user's non-stationary watching probability over the video progress bar, which additionally helps to
 respect the constraint of video length and facilitate the practical compatibility between the continuous regression event of watch time and other binary classification events.
We test our models extensively on two public datasets, a large-scale offline industrial dataset, and an online A/B test on a short video platform with hundreds of millions of daily-active users.
On all experiments, our models perform competitively against strong relevant baselines, demonstrating the efficacy of our user-centric perspective and proposed framework. 
\end{abstract}




\maketitle

\input{tex/intro}

\input{tex/related}

\input{tex/method}

\input{tex/experiment}
\input{tex/conclusion}


\begin{acks}
We’d like to thank Ali Selman Aydin,  Hongjie Bai,  Vladimir Batygin,  Jack Chai, Assaf Cohen,  Shilin Ding,  Yuhao Du, Bo Feng,  Gavin Feuer,  Chenzhang He,  Wenjie Hu,  Xinyao Hu,  Cheng Huang, Gong Ke,  Romaine Knight,  Chunlei Li,  Rui Li,  Sophia Liang,  Yijia Liu, Matt Ma,  Penny Pan,  Jing Qian,  Yisong Song,  Chris Steger,  Dayong Wang,  Meihong Wang,  Zellux Wang,  Zhe Wang,  Chuanqi Wei,  Yanhong Wu,  Hong Yan,  Lei Yuan,  Hui Zhang,  Lizhu Zhang,  Mengchee Zhang,  Qunshu Zhang,  and Yuting Zhang for their general support of the project.
\end{acks}

\bibliographystyle{ACM-Reference-Format}
\balance
\bibliography{sample-base}

\input{tex/appendix}

\end{document}

%% file: math_commands.tex
\usepackage{amsmath,amsfonts,bm,amsthm,mathrsfs,algorithm,algorithmic,url,xcolor,mathtools} 
\usepackage{booktabs,enumitem,multicol,caption,subcaption,pifont,tikz,listings,wrapfig,lipsum,circledsteps,arydshln,hyperref,multirow}
\usepackage{relsize,footnote,makecell,graphicx,microtype,thmtools}

\hypersetup{
    colorlinks,
    linkcolor={blue!50!black},
    citecolor={blue!50!black},
    urlcolor={red!50!black}
}

\theoremstyle{plain}
\newtheorem{theorem}{Theorem}[section]

\theoremstyle{definition}

\newtheorem{assumption}{Assumption}
\theoremstyle{remark}
\newtheorem{remark}[theorem]{Remark}
\newtheorem*{theorem*}{Theorem}

\def\eqref#1{(\ref{#1})}

\def\eqref#1{Eq.~(\ref{#1})}

\def\1{\bm{1}}

\DeclareMathAlphabet{\mathsfit}{\encodingdefault}{\sfdefault}{m}{sl}
\SetMathAlphabet{\mathsfit}{bold}{\encodingdefault}{\sfdefault}{bx}{n}

\def\gL{{\mathcal{L}}}

\newcommand{\E}{\mathbb{E}}

\newcommand{\R}{\mathbb{R}}

\DeclareMathOperator*{\argmax}{arg\,max}

\newcommand{\abs}[1]{\left\vert#1\right\rvert}

\newcommand{\cbr}[1]{\left\{#1\right\}}
\newcommand{\br}[1]{\left(#1\right)}

\newcommand{\ie}{\textit{i.e.}}
\newcommand{\eg}{\textit{e.g.}}

\definecolor{brickred}{rgb}{0.8, 0.25, 0.33}
\definecolor{darkspringgreen}{rgb}{0.09, 0.45, 0.27}
\definecolor{applegreen}{rgb}{0.55, 0.71, 0.0}
\definecolor{brightmaroon}{rgb}{0.76, 0.13, 0.28}
\definecolor{burgundy}{rgb}{0.5, 0.0, 0.13}

%% file: tex/intro.tex

\section{Introduction} \label{sec:intro}


The popularization of online video platforms, such as YouTube Shorts, TikTok, and Instagram Reels, has rendered the rising importance of (short) video recommendation as effective large-scale information filtering and entertainment systems
\citep{zhou2010impact,davidson2010youtube,tang2017popularity,liu2019user,liu2021concept,gao2022graph}.
On these platforms, video watch time emerges as a central feedback signal, which indicates users' interests and 
serves as a core optimization metric for companies' commercialization \citep{cai2023reinforcing,cai2023two}.
This metric is further highlighted by the relative inappropriateness of classical engagement signals such as click-through rate, since recommended videos on these platforms are typically (re-)played automatically and hence user's click action is erased.
To improve user engagement on the platform, accurate prediction of a user's watch time on a video has become a central task in the ranking stage, which nevertheless has  been challenging 
\citep{zhao2023uncovering,sun2024cread,bai2024labelcraft,lin2024conditional}.

Unlike conventional binary prediction tasks, such as estimating click-through rate or like rate, video watch time prediction is significantly more complex, making  current state-of-the-art methods still less satisfactory. 
The complexity is not solely due to its basis in continuous signal estimation; rather, it is significantly influenced by
the complex user-video interaction during video playback \citep{zhao2023breakingcursequalitysaturation}.
Concretely, users may exhibit different behavior modes in watching recommended videos.
For example, some users may keep ``jumping'' on the video progress bar.
Some others, however, may watch the video sequentially, but quit in interim (or till finishing). 
Further, users' interests/retention in the video, formulated as the probability of watching a video second, may be non-stationary and keep changing over the video progress bar.
Besides, each video has its own length, which naturally constraints video watch time and constitutes a layer of difficulty in modeling watch time \citep{zhan2022deconfounding,zheng2022dvr}.

Despite the importance and challenges, prior works on modeling video watch time tend to overlook the importance of understanding user behavior in video watching, and typically focus on relatively black-box mechanical enhancements of the classical regression and/or classification objectives.
By its continuous and scalar-valued nature, video watch time is classically modeled as an ordinary regression problem \citep{zhan2022deconfounding,tang2023counterfactual}, 
whose performance is however limited due to data outliers and skewed distribution \citep{sun2024cread}.
\citet{covington2016deep} convert this regression problem into learning the odds of the video click-through-rate in the  click-through classification, by using the observed video watch time as the per-sample weight.
\citet{zhan2022deconfounding} split training data evenly by video duration, and learn a regression model for each duration group for group-wise watch-time quantiles.
\citet{lin2023tree} break the regression problem of video watch time into a sequence of classification problems arranged in a tree structure, motivated by regression trees \citep{hastie2009elements}.
Though some of these methods have been applied to industrial systems, they still leave clear room for further improvement, particularly on
 suiting watch time prediction to user behavior and accounting for the complexity of user-video interaction.

In this paper, we for the first time take on a user-centric perspective to advance video watch time prediction, from which we propose a white-box statistical framework for modeling user's video watch time.
In particular, to factor in users' various behavior modes, we distill domain knowledge of user behavior into various behavioral assumptions of watching (short) videos. 
These behavioral assumptions are then directly translated into statistical models of video watch time.
To cope with user's non-stationary watching probability over the video progress bar, we bucketize video progress bar into buckets/bins and learn an individual watching probability for each bucket.
The flexibility of bucketization enables straightforward statistical models without sacrificing overall model capacity.
Bucketization can further pad videos to a unified maximum length but implicitly respect the constraint of each video's length, since short videos could have watching probabilities concentrated on lower-value buckets (\eg, 5-10s) and long videos on higher-value buckets (\eg, 95-100s).
With these considerations, our framework allows a unified principled statistical model on videos of all lengths.

We conduct extensive experiments to verify the efficacy of models derived from our framework, including two public datasets (CIKM16 Cup and KuaiRec \citep{gao2022kuairec}) and an offline large-scale industrial dataset.
In both settings, our models compete favorably against the current state-of-the-art approaches.
We further implement our proposed framework onto a real-world recommendation system on a popular short video platform with hundreds of millions of daily-active  users, and conduct online A/B test, where
our method brings significant online gain in increasing  users' video watch time on relevant video groups.

%% file: tex/related.tex
\section{Related Work}\label{sec:related_work}

\subsection{Video Watch Time Prediction}\label{sec:related_work_vvs_predict}
Video watch time prediction aims to predict a user's watch time  on a video, given user and video information \citep{sun2024cread}.
By the scalar nature of watch time, one can intuitively train a model by standard regression loss, which however suffers from data outliers, skewed distribution, and incompatibility with classification-based events \citep{sun2024cread}. 
\citet{covington2016deep} improve YouTube video recommendation by using ground-truth video watch time as the per-sample weight to enhance the logistic regression of predicting impressed videos.
A  watch time prediction can then be conveniently obtained by exponentiating the learned odds, which suits well to the long-tail nature of video-watch time.
This approach, however, is only valid under the assumption that the rate of viewed/impressed video is low, and is thus inappropriate to most real-world short video platforms where videos are played automatically and hence all  are viewed.
Continuing this research thread, TPM \citep{lin2023tree} constructs a tree-like structure to transform the task of video watch time prediction into a series of classification problems, for directing the given sample to its corresponding leaf node.
The predicted watch time is then obtained by a simple model within each leaf node, such as the average of training samples within that leaf.
\citet{sun2024cread} too predict video watch time through multiple classification problems that are constructed by their proposed discretization method for better balancing the learning and restoration errors. 
\citet{lin2024conditional} utilize quantile regression \citep{koenker2005quantile} to capture the watch time distribution, for providing more accurate and robust estimates.

Apart from enhancing loss functions and/or model structures, another line of works are developed to mitigate duration bias, where training data are identified as typically being imbalanced towards having more videos with longer duration.
\citet{wu2018beyond} define new metrics to measure the video quality.
D2Q \citep{zhan2022deconfounding} adopts backdoor adjustment \citep{pearl2009causality} to the confounding effect \citep{mickey1989impact} of video length, conveniently implemented by binning data based on video length. 
\citet{zhao2023uncovering} further correct model training for users' video exploration (noisy watching), in additional to the duration bias, so that the learned model can more accurately reflect the users' interests.
\citet{zheng2022dvr} introduce Watch Time Gain as a new unbiased evaluation metric for (short-)video recommendation and adopt adversarial learning for unbiased user preference modeling.

In this paper, we improve watch time modeling by directly facing  complex user-video interaction.
From a distinct user-centric perspective, we develop a statistical framework for modeling video watch time.
By translating domain-knowledge-based user behavior assumptions into statistical models and employing bucketization strategy, our framework can suit users' different behavior modes in video watching and varying interests in the video over its horizon.

\subsection{Classification for Regression}
There has been a long history of tackling regression problems from a classification lens, especially when the precise numerics of the target is less important.
A classical example of this approach is ordinal regression, which models the ordinal nature of the data \citep{mccullagh1980regression} and is typically extended from classical classification models, \eg, logistic regression \citep{mccullagh1980regression}, SVM \citep{shashua2002ranking}, and decision tree \citep{frank2001simple}.
Ordinal regression has been applied to various applications, including age estimation \citep{niu2016ordinal,beckham2017unimodal}, depth estimation \citep{fu2018deep,diaz2019soft}, and pose estimation \citep{hsu2018quatnet}.
Classification has also been used to model integer-valued data, where each distinct integer is treated as a different class \citep{waugh1995extending}.
In computer vision, modeling pixels as discrete values via classification rather than continuous values has been well adopted for its representational and training advantages \citep{van2016pixel,van2017neural,oord2018representation}.

As will be seen in \cref{sec:method}, in this paper, we are motivated by this long history of addressing regression via classification to use discrete probabilistic models for video watch time. 
These models are translated from users' behavior modes in video watching, admit efficient closed-form watch time estimators, and are compatible with the classification losses of other events in training industrial recommendation systems.

%% file: tex/method.tex
\section{The Proposed Method: \algname}\label{sec:method}

Our ultimate goal is to statistically model the random variable video watch time, which we denote as $T$, where $T$ is non-negative integer valued, \ie, $T \in \cbr{0,1,2,\ldots}$.
The observed value of $T$ is denoted as $t$.
We do not manually exert an upper bound on user's watch time $T$ by the video length, allowing replaying the video  indefinite times, \eg, the app automatically replays the video once ended.
As will be seen later in this section, our bucketization strategy allows  the user to have different watching probabilities at later replays, \eg, negligible probabilities at higher-value buckets if fewer/no replays. 


As discussed in \cref{sec:intro}, we statistically model video watch time  based on our distinct user-centric perspective. 
Concretely, to suit users' different behavior modes in watching (short) videos, we build our statistical models by directly translating different user behavior assumptions, which are distillation of our domain knowledge.
To relieve the modeling burden while allow aligning with the video length constraint, our statistical models operate on the entire video horizon: the (infinitely long) video progress bar counting in possibly indefinite time automatic replays, which conceptually pads out all possible time seconds in a video.
Further, considering the users' varying interests/retention intention over the video horizon, we bucketize the infinite video horizon onto smaller buckets, which allows intuitive statistical modeling on bucket-specific watching probabilities while ensures model capacity and the realistic finite-time video replays. 
In inference time, our models permit efficient closed-form point estimates of video watch time via the expectation of the specified statistical models.
Additional benefits of the bucketization strategy on stabilizing training objective is discussed in details in \cref{sec:connection_wlr}.



\begin{figure}[t]
         \centering
         \includegraphics[width=0.45\textwidth]{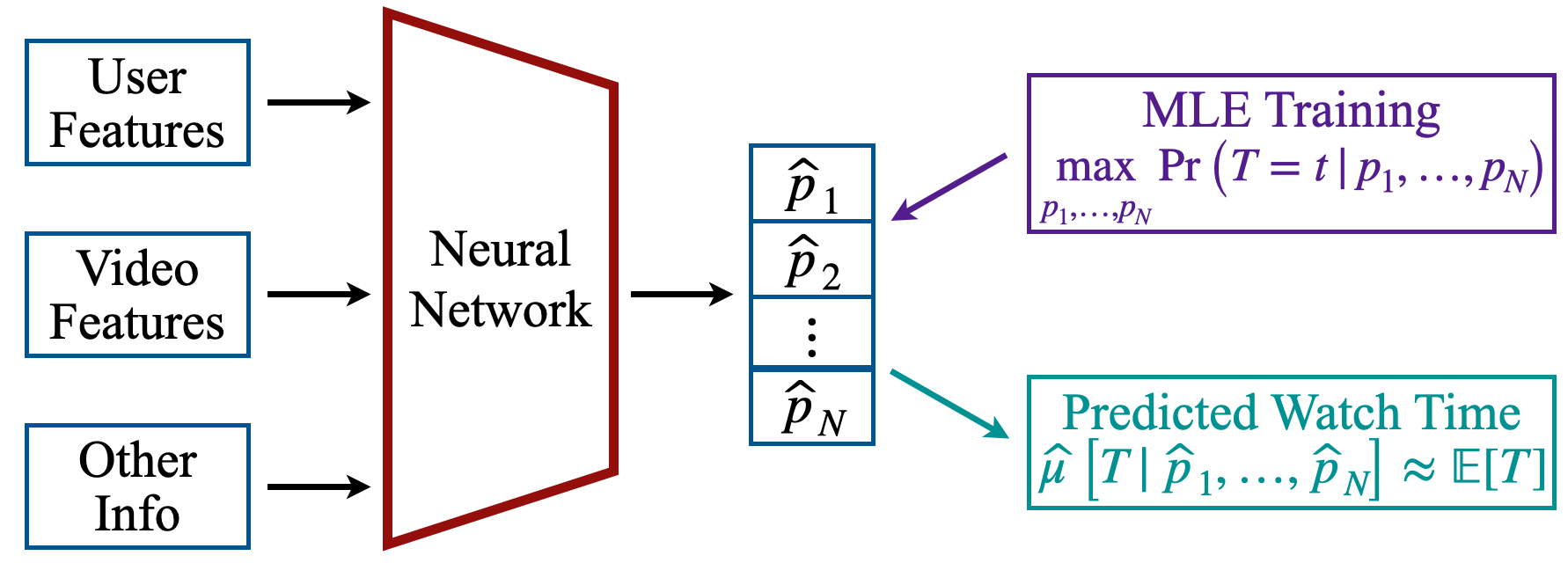}
         \caption{Overview of the {\algname} framework. The model outputs the watching probability $\widehat p_i$ for each bucket $B_i$, based on which an estimate $\widehat \mu[T]$ of the watch time $T$ is calculated.}
         \label{fig:arch}
         \vspace{-.5em}
\end{figure}

\subsection{Action Steps of \algname}\label{sec:method_general}
The first action step in our {\algname} framework is to partition the infinite video horizon $[0,\infty)$ by $N$ endpoints $\cbr{x_i}_{i=1}^N$, which facilitates specifying user behavior assumptions and learning non-stationary watching probabilities.
Notationally, we have: $-\epsilon=x_0 < x_1 < x_2 < \cdots < x_N < x_{N+1} = \infty$, where $-\epsilon \approx 0$ to conveniently include $t=0$ into the first bucket.
The $i^\mathrm{th}$ bucket is denoted as $B_i$ and the width of $B_i$ is $\Delta_i$, \ie, $B_i = (x_{i-1}, x_i], \Delta_i = x_i - x_{i-1}, \forall\, i=1, \ldots, N$.

With bucketization, different user behavior is factored in our statistical models by different behavioral assumptions that portray users' different behavior modes of watching (short) videos within each bucket.
These behavioral assumptions are then mapped to statistical models for video watch time, which depend on bucket-specific watching probability $p_i$ of each bucket $B_i$. 
Value variation of the learned $\cbr{\widehat p_i}$ manifests user's non-stationary interest/probability of continuing watching the video over its horizon.

With the observed video watch time $t$, our statistical models, in particular, $\{p_i\}$, are fitted by maximum likelihood estimation \citep[MLE,][]{ferguson1996,casellaberger2001}.
After fitting the model, in inference time, we estimate the expected watch time of a user on a video in closed-form based on the specified statistical model and the learned $\cbr{\widehat p_i}$, typically by summing up the user's predicted watch time in each bucket $B_i$, \ie, our estimate $\widehat \mu[T]$ of the expected video watch time $\E[T]$ is
\begin{equation*} \textstyle
   \E[T]\approx \widehat \mu[T] = \sum_{i=1}^{N} \text{``expected watch time in  $B_i$''}\;.
\end{equation*}
\cref{fig:arch} overviews our proposed framework.


In the next two sections, we will instantiate our general framework with two concrete yet distinct user behavior assumptions: a wandering-minded user (\cref{sec:bimon_model}) and a focused user (\cref{sec:geo_model}), based on which we derive principled statistical models and efficient watch time estimators.





\begin{figure}[t]
         \centering
         \includegraphics[width=0.47\textwidth]{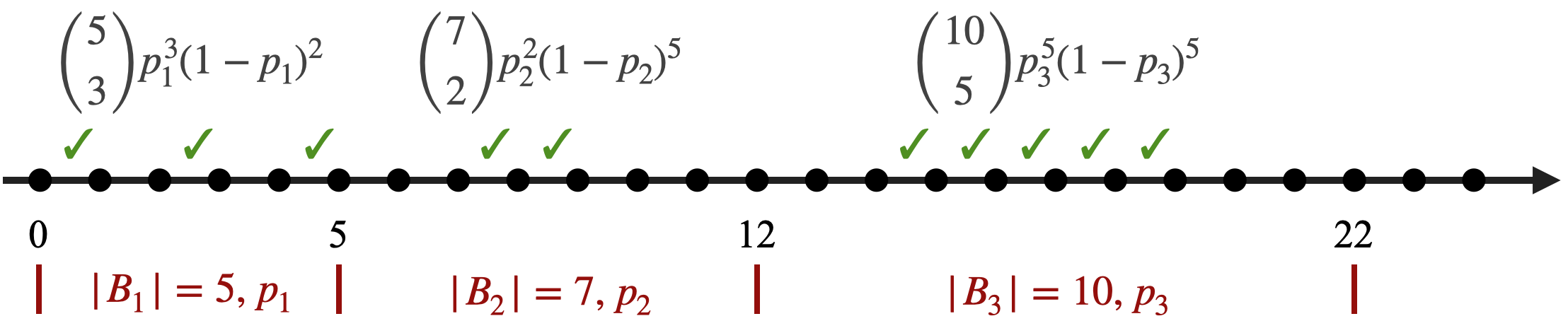}
         \caption{Illustration of the {\algname-Binom} model in \cref{sec:bimon_model}, showing three buckets respectively of length  $5$, $7$, and $10$. 
         The watch time in bucket 1 is $3$, bucket 2 is $2$, and bucket 3 is $5$.
         }
         \label{fig:binom_model}
         \vspace{-.5em}
\end{figure}

\subsection{{\algname-Binom}: Binomial Model for Random Pick-and-Play}\label{sec:bimon_model}
Our first behavior portray is a user with wandering minds: keeps jumping among buckets on the video horizon and within each bucket, only picks some (or zero) video seconds to watch.
\begin{assumption}\label{assump:binom_model}\label{assump:binom_model_jump_among_buckets}
    Within each bucket $B_i$, the user randomly picks some video seconds to watch, and the watching decision on each second is made independently with identical probability $p_i$.
    Furthermore, the user can freely move among buckets on the video horizon, without minimum watch time requirement at any bucket, mimicking jumping and/or skipping on the video progress bar.
\end{assumption}



    

Under Assumption~\ref{assump:binom_model}, in particular, identical watching probability in $B_i$ and  independence between watching the $i^{\mathrm{th}}$ second and the $(i+1)^{\mathrm{th}}$ second, the watch time in each bucket $B_i$ can be modeled as a binomial distribution. 
A pictorial illustration is in \cref{fig:binom_model}.
Technically, to avoid undesirable value truncation, $x_N$ should be chosen sufficiently large to cover the range of $T$ in practice.
The last bucket $[x_N, x_{N+1} = \infty)$ is not used due to its infinite support.
For an observation $t$ of $T$, denote the user's watch time in $B_i$ as $t_i$, where $\sum_{i=1}^N t_i = t$.
We use MLE to fit the watching probability $p_i$.
Concretely, for each $B_i$, with the given $\Delta_i$ and $t_i$, we have
\begin{equation}\label{eq:binom_likelihood_each_bin} \textstyle
\resizebox{0.44\textwidth}{!}{%
    $
    \begin{aligned}\textstyle
        &\argmax_{p_i}\; {\Delta_i \choose t_i} \, p_i^{t_i} \br{1-p_i}^{\Delta_i - t_i} \\
        \iff &  \argmax_{p_i}\;  p_i^{{t_i}/{\Delta_i}} \br{1-p_i}^{1 - {t_i}/{\Delta_i}}\\
        \iff & 
        \argmax_{p_i}\; \gL_{{\text{{\algname-Binom}}}} =: \frac{t_i}{\Delta_i}\log p_i + \br{1 - \frac{t_i}{\Delta_i}} \log\br{1-p_i}
        \,,
    \end{aligned}
$%
}    
\end{equation}
where the first equivalence is because the power function $(\cdot)^{1/\Delta_i}$ is strictly increasing on $\R^+$.
\cref{eq:binom_likelihood_each_bin} can be practically implemented by the classical binary cross-entropy loss with (soft) label $t_i/\Delta_i \in [0,1]$.
During model training, we optimize all $\cbr{p_i}_{i=1}^N$ jointly, while each $p_i$ has its own training objective. 
 Essentially, the neural network model is optimized for multiple classification problems simultaneously, implicitly enhancing  model training by multi-tasking \citep{caruana1997multitask}.

For the construction of (soft) label $t_i/\Delta_i$, since user's real per-bucket watch time $\cbr{t_i}_{i=1}^N$ is typically not recorded in current industrial recommendation systems.
We adopt an minimalist assumption that the \textit{recorded users} view the video stream sequentially, except in the last watched bucket, where the user gets bored and only picks some seconds to watch, and afterwards stops watching.
We note that this ``boredom'' assumption is not exerted in inference time, and is a practical compromise that can be removed by improved system logging.
With this extra assumption, we use the following construction of the label $t_i/\Delta_i$ based on the observation $t$: if $t \in B_n \iff x_{n-1} < t \leq x_n$, we have $t_i = \Delta_i, \forall\, i=1,\ldots, n-1$ and $t_i = 0, \forall\, i > n$.
The (soft) labels $\cbr{t_i/\Delta_i}_{i=1}^N$ are then constructed as,
\begin{equation}\label{eq:ple_label}
\resizebox{0.44\textwidth}{!}{%
    $
    \forall\, i = 1,\ldots, N,\; \text{$i^\mathrm{th}$ Soft Label}\; \frac{t_i}{\Delta_i} = \begin{cases}
        0 \,, & t \leq x_{i-1} \\ 
        1\,, & t > x_i \\
        \frac{t - x_{i-1}}{x_i - x_{i-1}}\,, & x_{i-1} < t \leq x_i
    \end{cases}\;.
$%
}  
\end{equation}

With the fitted $\widehat p_i$ for each bucket-specific watching probability $p_i$, we estimate expected watch time based on Assumption~\ref{assump:binom_model_jump_among_buckets} as
\begin{equation}\label{eq:binom_expect} \textstyle
    \widehat \mu_{\text{\algname-Binom}}[T] = \sum_{i=1}^N \Delta_i \times \widehat p_i \,,
\end{equation}
which is the sum of the expectation of binomial watch-second selection in each bucket.
Note that the user does not need to watch all/any seconds in prior buckets before jumping to next/later one.

\begin{remark}
    Even though the user may skip to later buckets as per Assumption~\ref{assump:binom_model_jump_among_buckets}, \cref{eq:ple_label} implicitly encourages the learned $\cbr{\widehat p_i}$ to be non-increasing, \ie, the probabilities of watching video seconds in later buckets is no greater than in earlier buckets, which implicitly respects the sequential mode of video watching.
\end{remark}

\begin{remark}
    \cref{eq:ple_label} is also used as piecewise linear encoding of an integer value feature for tabular deep learning \citep{gorishniy2022embeddings}, which has the benefit of allowing exact recovery of the feature value rather than the nearest bucket endpoint. 
\end{remark}

\begin{remark}
    Industrial recommendation systems are typically jointly trained by many events that are mostly classification-based, \eg, like and share.
    $\gL_{\text{\algname-Binom}}$ transforms the regression event of video watch time into (soft) classification, facilitating its compatibility with others in model training.
\end{remark}


\subsection{{\algname-Geo}: Geometric Model for Sequential  Watching}\label{sec:geo_model}

\subsubsection{Non-stationary Watching Probabilities over Video Horizon}\label{sec:bucket_geo_model}
Contrary to the wandering-minded user in \cref{sec:bimon_model},
our next behavior portray is a focused user: sequentially watches the video until a certain time point, when the user decides to stop.

\begin{assumption}\label{assump:bucket_geometric}
    At each second, the user decides whether to stop watching the video or  continue another second.
    The decision at each second is made independently, and within each bucket $B_i$, with identical probability $p_i$ for watching one more second.
\end{assumption}

\begin{remark}
Assumption~\ref{assump:bucket_geometric} accounts for user's varying interests/retention in the video by bucketizing the video horizon and prescribing a specialized watching probability $p_i$ for each bucket $B_i$.
\end{remark}

\begin{figure}[t]
         \centering
         \includegraphics[width=0.47\textwidth]{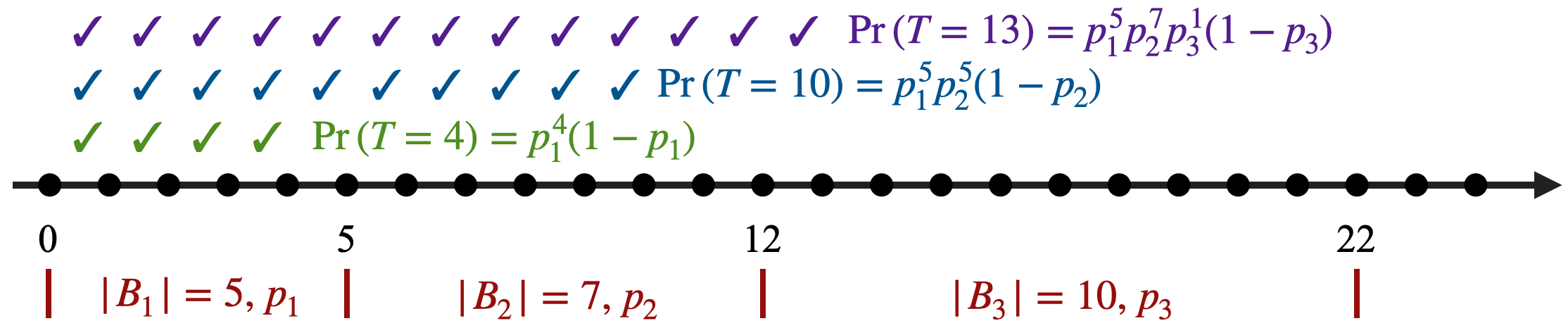}
         \caption{Illustration of the {\algname-Geo} model in \cref{sec:bucket_geo_model}, showing three buckets of lengths  $5$, $7$, and $10$ and the probability calculations of total watch time $T=4,10,13$.}
         \label{fig:b_geo_model}
\end{figure}

Under Assumption~\ref{assump:bucket_geometric}, user's video watch time can be modeled as a bucketized geometric distribution.
Specifically, for a given set of watching probabilities $\cbr{p_i}_{i=1}^{N+1}$, if the observation $t \in B_n = (x_{n-1}, x_n]$, the probability of $T=t$ is
\begin{equation}\label{eq:bucket_geometric_prob} \textstyle
    \Pr\br{T=t} = p_n^{t-x_{n-1}} \times (1-p_n)\times \prod_{i=1}^{n-1} p_i^{\Delta_i}\,,
\end{equation}
which also applies to the unbounded last bucket $(x_N, \infty)$.
A pictorial illustration of this model and the probability calculation is in \cref{fig:b_geo_model}.
An immediate benefit of this bucketized geometric model over the binomial model {\algname-Binom} in \cref{sec:bimon_model} is that \cref{eq:bucket_geometric_prob} does not require setting $x_N$ large enough to cover the unknown range of $T$, and will not have the otherwise undesirable value truncation.


\cref{eq:bucket_geometric_prob} allows us to fit the watching probabilities $\cbr{p_i}_{i=1}^{N+1}$ by maximum-likelihood estimation, leading to the objective
\begin{equation}\label{eq:bucket_geometric_ll} 
    \begin{aligned} \textstyle
        \argmax_{p_1,\ldots, p_{N+1}}\; \gL_{\mathrm{{\algname-Geo}}} &=:  \br{t-x_{n-1}}  \log p_n \,+ \\ &\quad\log \br{1-p_n} + \sum_{i=1}^{n-1} \Delta_i \log p_i
        \,,
    \end{aligned}
\end{equation}
where the index $n$ refers to the bucket $B_n$ that the given training sample falls into and may differ across samples. 

With predicted $\widehat p_i$ for each bucket-specific watching probability $p_i$, we estimate the user's expected watch time on a given video as
\begin{equation}\label{eq:bucket_geo_expect}
\resizebox{0.44\textwidth}{!}{%
    $
    \widehat \mu_{\text{\algname-Geo}}[T] = \sum_{i=1}^{N+1} \prod_{j=1}^{i-1} \widehat p_j^{\Delta_j} \br{x_{i-1}\widehat p_i + \frac{\widehat p_i \br{1-\widehat p_i^{\Delta_i}}}{1-\widehat p_i} - x_i\widehat p_i^{\Delta_i + 1}} \,,
$%
} 
\end{equation}
with notational assumptions that $\widehat p_0 = 1, p_{N+1}^\infty = 0, x_{N+1}\widehat p_{N+1}^{\infty} = \infty \widehat p_{N+1}^{\infty} = 0$ for equation simplicity.
Detailed proof of \cref{eq:bucket_geo_expect} is deferred to \cref{sec:proof_exp_bucket_geo}.

\subsubsection{Simplification: A Stationary Watching Probability}\label{sec:vanilla_geo_model}

One may simplify the assumption on non-stationary watching probabilities in \cref{sec:bucket_geo_model} by assuming a uniform watching probability over the entire video horizon, \ie, the user's interest in continuing watching the video does not vary with the played time.
As will be revealed later on and in \cref{sec:connection_wlr}, this simplified model is closely related to weighted logistic regression \citep[WLR,][]{covington2016deep}.

\begin{assumption}\label{assump:vanilla_geometric}
    At each second, the user decides whether to stop watching the video or to continue another second.
    The decision at each second is made independently, and over the entire video horizon, with identical  probability $p$ for watching one more second.
\end{assumption}

Assumption~\ref{assump:vanilla_geometric} translates to the vanilla geometric distribution for modeling video watch time, where $p$ is interpreted as the ``success probability'' for the event of ``watching another second''.
The probability mass function of this model and expectation of video watch time is, $\forall\,t \in \cbr{0,1,2,\ldots}$,
\begin{equation}\label{eq:vanilla_geo_mean} \textstyle
    \begin{aligned}
       \Pr\br{T=t} = p^t (1-p) \,, \;
       \E[T] = p / (1-p) = 1/(1-p) - 1 
        \,.
    \end{aligned}
\end{equation}
The objective for maximum-likelihood fitting of $p$ is 
\begin{equation}\label{eq:vanilla_geo_ll}
    \begin{aligned} \textstyle
       \argmax_{p}\; \gL_{\mathrm{v-geo}} =: t\log p + \log(1-p)\,,\; \forall\, t \in \mathbb{N} \,,
    \end{aligned}
\end{equation}
where the observed watch time $t$ is again given for training samples.

If the estimate $\widehat p$ is parameterized by the sigmoid function $\widehat p = 1/ (1+\exp(-y))$ where $y$ is the unnormalized logit from the neural network.
The estimate of expected watch time admits a simple form
\begin{equation}\label{eq:vanilla_geo_expect}
\resizebox{0.44\textwidth}{!}{%
    $
    \widehat \mu_{\mathrm{v-geo}}[T] = \frac{1}{1-\widehat p} - 1 = \frac{1}{\frac{\exp(-y)}{1+\exp(-y)}} - 1 = e^y + 1 -1 = e^y  \,.
$%
} 
\end{equation}

\begin{remark}\label{remark:wlr_inference_time}
    While \cref{eq:vanilla_geo_expect} takes the same functional form as the watch-time estimator in WLR \citep{covington2016deep}, it operates  without extra assumption or approximation in our framework.
    Concretely, \citet{covington2016deep} also estimate watch time by the exponentiated logits (\cref{eq:vanilla_geo_expect}),
    but require the extra assumption of \textit{small click probability} to establish certain approximation. Click probability is inapproprate to our target of short video platforms, as discussed in Sec. \ref{sec:intro} and \ref{sec:related_work_vvs_predict}.
\end{remark}

\subsection{Connection with WLR}\label{sec:connection_wlr}

Vanilla geometric model \cref{eq:vanilla_geo_ll} can be interpreted as a regularized, yet statistically principled, extension of WLR.
Concretely, the log-likelihood objective in WLR and in vanilla geometric model is,
\begin{equation}\label{eq:ll_wlr}
\resizebox{0.43\textwidth}{!}{%
    $
    \gL_{\mathrm{WLR}} =: \begin{cases}
        t \log p\,,  t > 0 \\
        \log(1-p)\,,  t = 0
    \end{cases} v.s.\,
    \gL_{\mathrm{v-geo}} = \begin{cases}
        t  \log p \textcolor{burgundy}{+ \log(1-p)} \,,  t > 0 \\
        \log(1-p)\,,  t = 0
    \end{cases}
$%
} 
\end{equation}
In \cref{eq:ll_wlr}, $\gL_{\mathrm{WLR}}$ and $\gL_{\mathrm{v-geo}}$ have the same form when $t=0$.
However, when $t>0$, $\gL_{\mathrm{v-geo}}$ regularizes the fitting of $p$ with a regularization term $\log(1-p)$, which helps to mitigate overly large and/or highly varying gradient to increase $p$ when $t$ is large/varying and the resulting training instability and overfitting tendency. 

Though incorporating a regularization term compared to WLR, $\gL_{\mathrm{v-geo}}$ still has its objective value scale with the unbounded observed watch time $t$.
The bucketized geometric objective $\gL_{\text{{\algname-Geo}}}$ in  \cref{eq:bucket_geometric_ll} further stabilizes training via bucket-specific  watching probabilities $\cbr{p_i}$: each $\log p_i$ term in the  objective \cref{eq:bucket_geometric_ll} has weight no greater than the bounded bucket width $\Delta_i$, except the rare case where training sample falls into the last bucket $B_{N+1}$ ($p_{N+1}$), which will nevertheless have a regularized weight $t-x_N \ll t$. 

Further, with bucketization,  the binomial objective 
$\gL_{\text{{\algname-Binom}}}$ in \eqref{eq:binom_likelihood_each_bin}  also has similar benefits in regularizing objective value and thus training stability.
Concretely, plugging the  soft label constructed in \eqref{eq:ple_label} back to the training objective \eqref{eq:binom_likelihood_each_bin}, we have
\begin{equation}\label{eq:binom_likelihood_expand} \textstyle
\resizebox{0.43\textwidth}{!}{%
    $
    \gL_{\text{{\algname-Binom}}} = \begin{cases}
        \log(1-p_i)\,, & t \leq x_{i-1} \\
        \log p_i \,, & t > x_i \\
        \frac{t - x_{i-1}}{x_i - x_{i-1}} \log p_i + \frac{x_i - t}{x_i - x_{i-1}} \log(1-p_i) \,, & x_{i-1} < t \leq x_i
    \end{cases}
$%
}    
\end{equation}
From \cref{eq:binom_likelihood_expand}, it is clear that in jointly optimizing the neural network model for learning $\cbr{p_i}_{i=1}^N$, each $\log p_i$ and/or $\log (1-p_i)$ term in the objective \cref{eq:binom_likelihood_expand} has weight at most $1$.
Hence $\gL_{\text{{\algname-Binom}}}$ too avoids the unbounded weight in $\gL_{\mathrm{WLR}}$ and $\gL_{\mathrm{v-geo}}$ shown in \cref{eq:ll_wlr} (\ie, observed watch time $t$ itself) and can also be viewed as a principled weight-stabilized improvement on WLR.


As a final remark, the functional similarity between $\gL_{\mathrm{WLR}}$ and $\gL_{\mathrm{v-geo}}$ indicates that WLR approximately models video watch time by a geometric distribution with \textit{uniform} watching probability over the entire video horizon, which accounts 
for \textit{neither} real user's varying interests in continuing watching  \textit{nor} different video lengths.
By contrast, the bucketization strategy employed in our framework can naturally cope with both desiderata should data support.
For example, to account for the differences in video length, the model may learn   unequal $\{\widehat p_i\}$ whose values become  negligible  when the index $i$ is greater than some smaller integers for short videos (due to their smaller lengths) and larger integers for long videos.


%% file: tex/experiment.tex
\section{Experiments}\label{sec:exp}
We conduct extensive experiments of various nature to testify the efficacy of  models derived from our proposed framework in real-world scenarios.
Concretely, \cref{sec:public_data} presents the evaluation of our {\algname-Binom} model (\cref{sec:bimon_model}) and {\algname-Geo} model (\cref{sec:bucket_geo_model}) on two public recommendation datasets.
\cref{sec:offline_exp} evaluates our models on a large-scale industrial dataset.
In \cref{sec:online_exp}, we conduct an online A/B test
by implementing our methods onto a real-world recommendation system on a popular short video platform with hundreds of millions of daily-active users. 


\subsection{Public Dataset}\label{sec:public_data}
\subsubsection{Datasets.}
We follow \citet{lin2023tree} to test our models on two public datasets CIKM16 Cup and KuaiRec \citep{gao2022kuairec}.

CIKM16 Cup dataset contains sequences of anonymous transactions provided by DIGINETICA, and the goal is to predict the dwell time in search results for each session.
The dataset contains $310,302$ sessions, a total of $122,991$ items and the average dwell time of each session is $3.981$.
For each session, each item in this session is used as a feature in the model input.

KuaiRec dataset is collected from the recommendation logs of the short video platform Kuaishou and is one of the few datasets that provide a dense user-item interaction matrix, \ie, each user has watched each video and left feedback.
KuaiRec dataset contains $7,176$ users, $10,728$ items, and $12,530,806$ interactions.
As in \citet{lin2023tree}, for both public datasets, training and testing set are randomly split according to $80-20\%$ ratio.

\subsubsection{Baselines.}
We compare our method with state-of-the-art methods for video watch time prediction, including Weighted Logistic Regression \citep[WLR,][]{covington2016deep}, Duration-Deconfounded Quantile Regression \citep[D2Q,][]{zhan2022deconfounding}, Ordinal Regression \citep[OR,][]{crammer2001pranking}, Tree-based Progressive Regression Model \citep[TPM,][]{lin2023tree}, and Classification-Restoration framework with Error Adaptive Discretization \citep[CREAD,][]{sun2024cread}.
A detailed description of these baseline methods is provided in \cref{sec:details_public_dataset}.

\subsubsection{Implementation Details.}\label{sec:public_data_imple_details}
To ensure fair comparisons, the implementation of our models in this section strictly follows the release source code of \citet{lin2023tree},  except the  loss function and the corresponding model output layer.
We inherit as many hyperparameter settings as possible from \citet{lin2023tree}.
The models are trained until the training losses converge.
Note that geometric distribution is only validly defined on integers, while the target values of both KuaiRec and CIKM16 datasets are real scalars. 
For the {\algname-Geo} model, we therefore construct the training target as
\begin{equation*}
    \text{training target} = \mathrm{round}(c \times \text{data target value})\,,
\end{equation*}
where $\mathrm{round}(\cdot)$ denotes rounding to the nearest integer, $c=50$ for the KuaiRec dataset, and $c=100$ for the CIKM dataset, both of which are conveniently chosen based on the scale of the data target values.
In inference time, the effect of multiplication by $c$ is reverted before calculating the evaluation metrics.
On both datasets, we conveniently use $100$ buckets, defined by the $1$-percentiles of the data target values.
\cref{sec:exp_public_data_ablation} conducts ablation study on the number of buckets, showing that our method is generally robust to this design parameter.
On the CIKM dataset, both {\algname-Binom} and {\algname-Geo} use learning rate 2e-3. On the KuaiRec dataset, {\algname-Binom} uses learning rate 3e-3 and {\algname-Geo} uses 5e-4. These learning rates were selected simply by the converged value of the \textit{training loss} and the speed of convergence.
We use AdamW optimizer \citep{adamw2017} with weight decay $0$.

\subsubsection{Evaluation Metrics.}
We evaluate our models' prediction in terms of both numerical accuracy and ranking capability, the latter of which is important for selecting videos to recommend.
We hence follow \citet{lin2023tree} to evaluate our models on two metrics Mean Absolute Error (MAE) and XAUC \citep{zhan2022deconfounding}, which are defined as follows.
\begin{itemize}[leftmargin=*]
\item \textit{MAE}: This metric measures absolute numerical error.
Denote the observed target value as $t_i$ and the predicted value as $\widehat t_i$, we have
\begin{equation*}\textstyle
    \text{MAE} = \frac{1}{N}\sum_{i=1}^N \abs{\widehat t_i - t_i} \,,
\end{equation*}
where $N$ denotes the test set size.
Smaller MAE suggests better model performance.
\item \textit{XAUC}: This metric is an extension of AUC to the scenario where both prediction and target are real valued, as in the case of video watch time. 
For each pair of samples, a score of $1$ is assigned if the order of the predicted values of these two samples are the same as the order of the ground truth target values, otherwise a score of $0$ is assigned.
Sample pairs are uniformly drawn from the test set and the averaged score is reported, which measures the percentage of samples where the pairwise ordering is correctly predicted.
Larger XAUC indicates better model performance.
\end{itemize}
As a side note, we observed that our models' performance is stable and roughly the same across multiple random seeds. 
For better comparison with baseline methods, we followed \citet{lin2023tree} and \citet{sun2024cread} to not include standard deviations in the result table.

\begin{table}[t]
\caption{Performance of our models on the metrics XAUC and MAE on KuaiRec and CIKM16 datasets.
Baseline results are directed cited from \citet{lin2023tree} and \citet{sun2024cread}.
The best result in each metric is \textbf{bold}, and the second best result is \textbf{\textit{bold italic}}. 
}
\label{table:public_main}
\setlength\tabcolsep{6pt}
\vspace{-0.5em}
\begin{tabular}{@{}ccccccc@{}}
\toprule
\multirow{2}{*}{Method} & & \multicolumn{2}{c}{KuaiRec}     &  & \multicolumn{2}{c}{CIKM16}      \\ \cmidrule(lr){3-4} \cmidrule(l){6-7} 
                  &      & MAE $\downarrow$            & XAUC   $\uparrow$        &  & MAE  $\downarrow$          & XAUC   $\uparrow$        \\ \midrule
WLR       &              & 6.047          & 0.525          &  & 0.998          & 0.672          \\
D2Q      &               & 5.426          & 0.565          &  & 0.899          & 0.661          \\
OR       &               & 5.321          & 0.558          &  & 0.918          & 0.664          \\
TPM        &             & 4.741          & 0.599          &  & 0.884          & 0.676          \\
CREAD        &             & \textbf{3.215}          & 0.601          &  & 1.793          & 0.667          \\ \midrule
\textbf{{\algname-Binom}}   &                & \textbf{\textit{3.363}} & \textbf{0.609} &  & \textbf{0.847} & \textbf{\textit{0.683}} \\
\textbf{{\algname-Geo}}    &               & {3.589} & \textbf{\textit{0.603}} &  & \textbf{\textit{0.864}} & \textbf{0.688} \\ \bottomrule
\end{tabular}
\end{table}


\begin{figure*}[tb]
     \centering
     \begin{subfigure}[b]{0.24\textwidth}
         \centering
         \includegraphics[width=\textwidth]{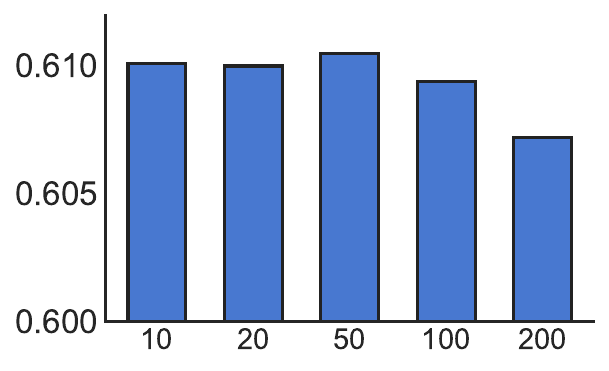}
         \caption{{\algname-Binom} - XAUC $\uparrow$} 
         \label{fig:kuai_binom_xauc}
     \end{subfigure}
     \hfill
     \begin{subfigure}[b]{0.24\textwidth}
         \centering
         \includegraphics[width=\textwidth]{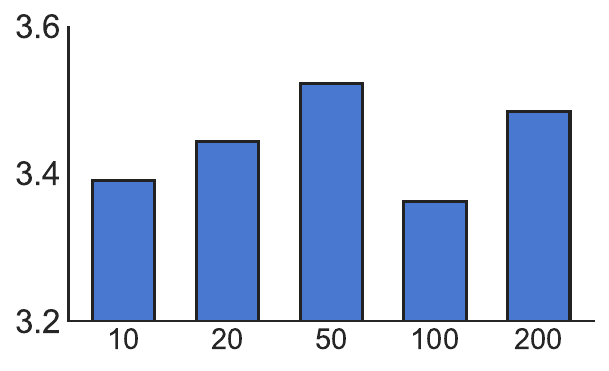}
         \caption{{\algname-Binom} - MAE $\downarrow$} 
         \label{fig:kuai_binom_mae}
     \end{subfigure}
     \hfill
    \begin{subfigure}[b]{0.24\textwidth}
         \centering
         \includegraphics[width=\textwidth]{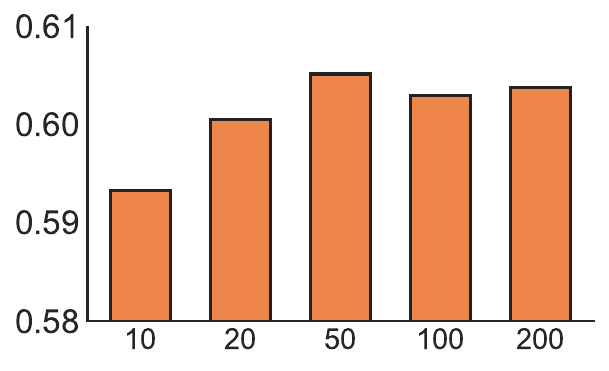}
         \caption{{\algname-Geo} - XAUC $\uparrow$}    
         \label{fig:kuai_geo_xauc}
     \end{subfigure}
     \hfill
     \begin{subfigure}[b]{0.24\textwidth}
         \centering
         \includegraphics[width=\textwidth]{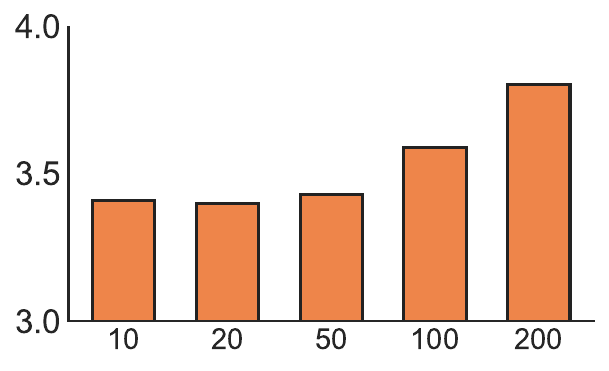}
         \caption{{\algname-Geo} - MAE $\downarrow$} 
         \label{fig:kuai_geo_mae}
     \end{subfigure}
        \caption{ 
        Performance of the {\algname-Binom} model (\cref{sec:bimon_model}) and {\algname-Geo} model (\cref{sec:bucket_geo_model}) on the KuaiRec dataset on the metrics XAUC and MAE, when varying the number of buckets across $\cbr{10,20,50,100,200}$.
        Horizontal axis denotes the number of buckets and vertical axis denotes the metric value.
        }
        \label{fig:kuai_sweep_bins}
\end{figure*}

\begin{figure*}[tb]
     \centering
     \begin{subfigure}[b]{0.24\textwidth}
         \centering
         \includegraphics[width=\textwidth]{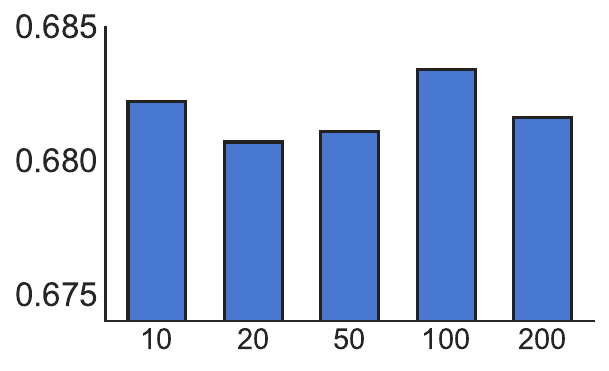}
         \caption{{\algname-Binom} - XAUC $\uparrow$} 
         \label{fig:cikm_binom_xauc}
     \end{subfigure}
     \hfill
     \begin{subfigure}[b]{0.24\textwidth}
         \centering
         \includegraphics[width=\textwidth]{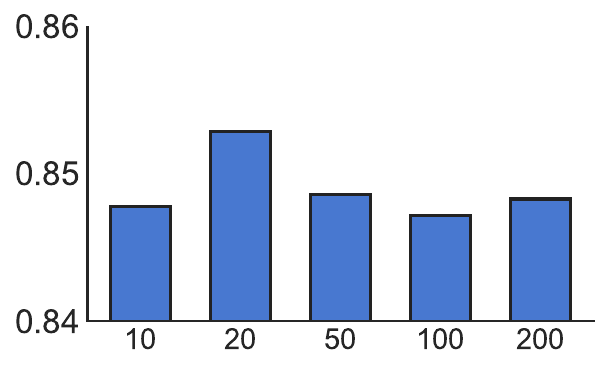}
         \caption{{\algname-Binom} - MAE $\downarrow$} 
         \label{fig:cikm_binom_mae}
     \end{subfigure}
     \hfill
    \begin{subfigure}[b]{0.24\textwidth}
         \centering
         \includegraphics[width=\textwidth]{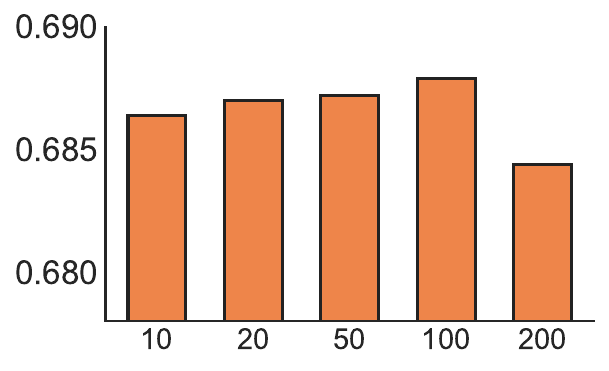}
         \caption{{\algname-Geo} - XAUC $\uparrow$}    
         \label{fig:cikm_geo_xauc}
     \end{subfigure}
     \hfill
     \begin{subfigure}[b]{0.24\textwidth}
         \centering
         \includegraphics[width=\textwidth]{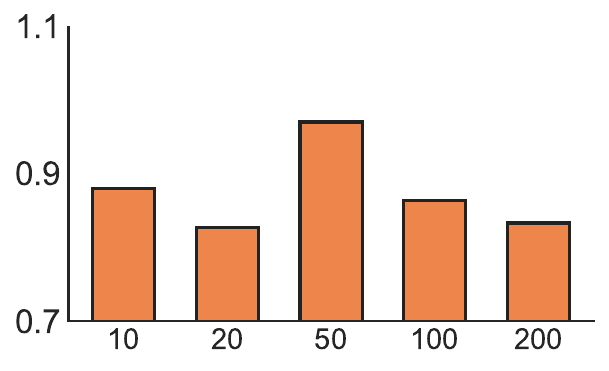}
         \caption{{\algname-Geo} - MAE $\downarrow$} 
         \label{fig:cikm_geo_mae}
     \end{subfigure}
        \caption{ 
        Performance of the {\algname-Binom} model (\cref{sec:bimon_model}) and {\algname-Geo} model (\cref{sec:bucket_geo_model}) on the CIKM16 dataset on the metrics XAUC and MAE, when varying the number of buckets across $\cbr{10,20,50,100,200}$.
        Horizontal axis denotes the number of buckets and vertical axis denotes the metric value.
        }
        \label{fig:cikm_sweep_bins}
\end{figure*}

\begin{figure*}[tb]
    \centering
\includegraphics[width=0.45\textwidth]{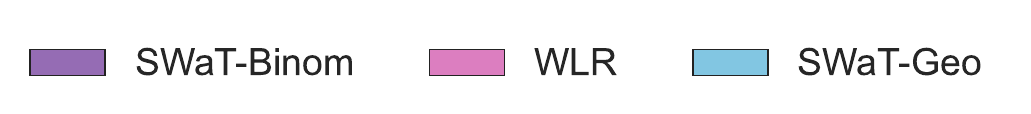}
\\
     \begin{subfigure}[b]{0.24\textwidth}
         \centering
         \includegraphics[width=\textwidth]{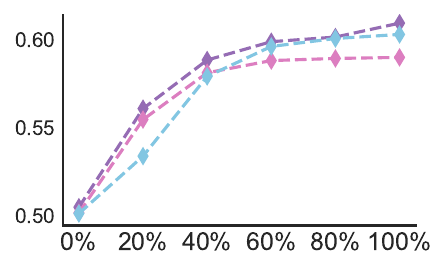}
         \caption{KuaiRec - XAUC $\uparrow$} 
         \label{fig:kuai_xauc_line}
     \end{subfigure}
     \hfill
     \begin{subfigure}[b]{0.24\textwidth}
         \centering
         \includegraphics[width=\textwidth]{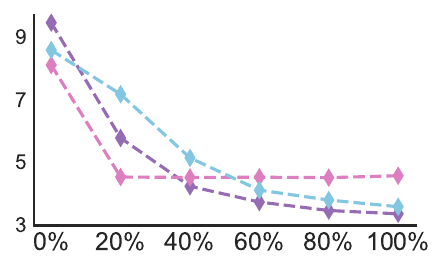}
         \caption{KuaiRec - MAE $\downarrow$} 
         \label{fig:kuai_mae_line}
     \end{subfigure}
     \hfill
     \begin{subfigure}[b]{0.24\textwidth}
         \centering
         \includegraphics[width=\textwidth]{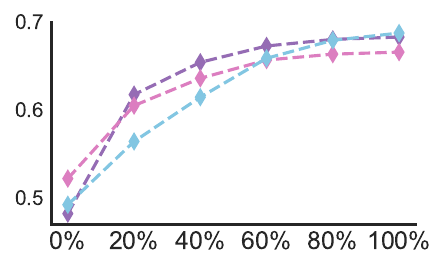}
         \caption{CIKM16 - XAUC $\uparrow$} 
         \label{fig:cikm_xauc_Line}
     \end{subfigure}
     \hfill
     \begin{subfigure}[b]{0.24\textwidth}
         \centering
         \includegraphics[width=\textwidth]{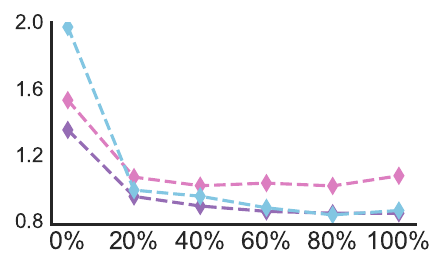}
         \caption{CIKM16 - MAE $\downarrow$} 
         \label{fig:cikm_mae_line}
     \end{subfigure}
        \caption{ 
         Performance on the metrics XAUC and MAE over the training process, comparing our {\algname-Binom} model (\cref{sec:bimon_model}) and {\algname-Geo} model (\cref{sec:bucket_geo_model}) with WLR \citep{covington2016deep} on both KuaiRec and CIKM16 dataset. 
        Horizontal axis represents $t\%$ of the training process.
        The curves for WLR are from our re-implementation based on \citet{zhan2022deconfounding} and can deviate from the results in \cref{table:public_main}, which we cite from \citet{lin2023tree}.
        }
        \label{fig:line_training_curve}
\end{figure*}

\subsubsection{Main Results.}
We compare the results of our models and the baselines in \cref{table:public_main}, where the baseline results are directly cited from \citet{lin2023tree} and \citet{sun2024cread}.
In \cref{table:public_main}, our two models {\algname-Binom} and {\algname-Geo} perform stably across all the metrics and datasets and are almost always ranked top-2.
The better performance of our models over the SoTA methods TPM and CREAD indicates the efficacy of our user-centric perspective and statistical framework for modeling video watch time.
The gain of our methods over WLR may further testify our consideration on training stability and regularization, which are naturally incorporated in our proposed framework, as discussed in \cref{sec:connection_wlr}. 

\subsubsection{Ablation Study.}\label{sec:exp_public_data_ablation}

This section discusses the following two research questions to better verify our method.

\textbf{(a):} \textit{Are our models robust to the number of buckets?}

As discussed in \cref{sec:public_data_imple_details}, in the public dataset experiments, we define buckets by the percentiles of the data target values.
Therefore, the only hyperparmeter introduced by our {\algname-Binom} and {\algname-Geo} models is the number of buckets.
We hence vary the number of buckets across $\cbr{10,20,50,100,200}$, respectively defining the buckets as the $\cbr{10, 5, 2, 1, 0.5}$-percentiles of the data target values.
\cref{fig:kuai_sweep_bins} shows the performance of our {\algname-Binom} and {\algname-Geo} models on  the metrics XAUC and MAE on the KuaiRec dataset.
\cref{fig:cikm_sweep_bins} plots the results on the CIKM16 Cup dataset.
We encourage the reader to review the best baseline results in \cref{table:public_main}, which are omitted in these two figures for better visualization. 

From \cref{fig:kuai_sweep_bins} and \cref{fig:cikm_sweep_bins} and recall the best baseline results in \cref{table:public_main}, it is clear that our models are generally robust to the number of buckets.
While our convenient choice of $100$ buckets in \cref{sec:public_data_imple_details} works well across the combination of models, metrics, and datasets, performance on a specific setting may  be further improved by tuning the bucket numbers.
For example, on the KuaiRec dataset, $50$ buckets suits {\algname-Geo} model best, and using $20$ or even $10$ buckets can improve {\algname-Binom} model's XAUC.
Using more buckets, such as $200$, however, generally does not provide much gain over using $100$ or fewer buckets.
Hence, for a practical rule-of-thumb, one may start with $100$ buckets, and gradually reduce the bucket number while closely monitoring the desired metrics.

\textbf{(b):} \textit{Is the training of our models stable and well converged?}

Given the complexity introduced by the bucketization process, one may naturally question the training stability of our models.
To verify  training stability and convergence, we  evaluate our models at $\cbr{0, 20,40,60,80,100}\%$ of the training process, in terms of the metrics XAUC and MAE.
 For better comparison, we also re-implement the baseline Weight Logistic Regression (WLR) based on \citet{zhan2022deconfounding} to plot its training curves.
Similar to our models, WLR is trained until the training loss converges and its (training) hyperparameters are carefully tuned.
Note that the results of WLR here can deviate from our cited results in \cref{table:public_main} due to various uncontrollable issues, such as package version.
\cref{fig:line_training_curve} plots the training curve comparison on both KuaiRec and CIKM16 Cup datasets.

From \cref{fig:line_training_curve}, it is evident that our models converge stably and well across the combination of metrics and datasets.
In particular, the performance of our models keep improving over the training process, in drastic contrast to the training curves of WLR, which seem to converge too early on the first half of the training processes, especially on the KuaiRec dataset.
This may be caused by the training/gradient instability of the WLR method that we identify in \cref{sec:connection_wlr}: the loss value and thus gradient norm of WLR are directly determined by the highly varying data target values.
This issue is mitigated in our models through the bucketization strategy employed in our proposed statistical framework.

\subsection{Industrial Offline Experiment}\label{sec:offline_exp}
We validate the efficacy of our models on a large-scale industrial dataset collected from a popular short video platform.

\subsubsection{Datasets.}
In the industrial offline evaluation, we train the model by using user interactions on the platform in $3$ days  and test the models' performance on  the following day's data.
The training set contains more than $5 \times 10^{10}$ samples and the evaluation set contains more than $1.5\times 10^{10}$ samples.

\subsubsection{Baseline.}\label{sec:offline_baseline}
The baseline model is trained by the classical binary cross-entropy loss for predicting video watch time, which is similar to our {\algname-Binom} model (\cref{sec:bimon_model}) but \textit{without} bucketization and bucket-specific watching probabilities.

\subsubsection{Implementation Details.}
Due to the large size of the dataset, we train the models with one pass of the training set, \ie, for one epoch.
Our models and the baseline use the same model architecture, except for the inevitable changes in the loss function and the corresponding model output layer. 
All models use the same set of  features and hyperparameters, both of which are optimized for the baseline model, which leaves room for further improving our models' performance.
Bucket definition is investigated in \cref{sec:offline_abla}.

\subsubsection{Evaluation Metric.}
As per industrial common practice, we evaluate the prediction of video watch time by the Pearson correlation between the predicted watch time and the ground truth.
Large Pearson correlation indicates that higher predicted values mostly correspond to higher ground-truth values, and thus the prediction model can be more helpful for selecting videos to recommend.
As an aside, we re-ran our models several times using data from different time periods, and the performance gains indicated in \cref{table:offline_main} were robust.

\subsubsection{Main Results}

\begin{table}[]
\caption{
Performance of our models and baseline model on the industrial offline dataset.
``V-Geo'' denotes the vanilla geometric model in \cref{sec:vanilla_geo_model}.
}
\label{table:offline_main}
\setlength\tabcolsep{3pt}
\vspace{-0.5em}
\begin{tabular}{@{}ccccc@{}}
\toprule
Model               & Baseline & {\algname-Binom} & {\algname-Geo} & V-Geo \\ \midrule
Pearson Corr. & 56.75\%  & 57.55\%  & 57.56\%    & 57.54\%     \\
Gain over Baseline  & -        & +0.80\%   & +0.81\%     & +0.79\%      \\ \bottomrule
\end{tabular}
\end{table}

\cref{table:offline_main} compares the performance of our three models discussed in \cref{sec:method} against the baseline model.
It is clear that all our proposals, even the un-bucketized vanilla geometric model, perform significantly better than the baseline, demonstrating the efficacy of our user-centric perspective and statistical framework for modeling video watch time.
As a side note, the simplest vanilla geometric model is less performant on other industrial topline metrics, and thus we will continue focusing on our two bucketized models.

\subsubsection{Ablation Study on the Bucket Definition.}\label{sec:offline_abla}

We investigate the performance of our {\algname-Binom} model and {\algname-Geo} model under various bucket definitions.
Our intention is to find a bucket definition that performs well while having fewer buckets.
We hypothesize that due to the long-tail distribution of video watch time and our desiderata of modeling the right tail well, a dense uniform grid may not be desirable. 
With this motivation, we compare the following six choices of the bucket endpoints.
\begin{itemize}
    \item[Choice 1] Uniform grid of the 5-percentiles of  video watch time values in the training set.
    \item[Choice 2] Uniform grid of the 2-percentiles of  video watch time values in the training set.
    \item[Choice 3] Uniform grid of the 1-percentiles of  video watch time values in the training set.
    \item[Choice 4] Uniform grid of the 2-percentiles of video watch time values until the 96 percentile, concatenated by the grid of the 5-percentiles of the remaining 4\% of data.
    \item[Choice 5]  
    Uniform grid of the 2-percentiles of  video watch time values until the 96 percentile, concatenated by the grid of the 2-percentiles of the remaining 4\% of data.
    \item[Choice 6]  
    Uniform grid of the 1-percentiles of  video watch time values until the 90 percentile, concatenated by the grid of the 1-percentiles of the remaining 10\% of data.
\end{itemize}
Choices 4-6 allow a finer grid for large video watch time values, \ie, the right tail of the data distribution.

\begin{table}[]
\caption{{\algname-Binom} and {\algname-Geo} model's \textit{gain over the baseline model}, under different choices of bucket endpoints  as listed in \cref{sec:offline_abla}, together with the number of buckets under each choice.
The gain is calculated as a direct subtraction by the baseline model's result in \cref{table:offline_main}.
Highest numeric in each column is bold.
}
\label{table:bucket_selection}
\setlength\tabcolsep{6pt}
\vspace{-0.5em}
\begin{tabular}{@{}lccc@{}}
\toprule
         & {\algname-Binom}     & {\algname-Geo} & Num. Buckets \\ \midrule
Choice 1 & 0.71\%        & 0.44\%                 & 18           \\
Choice 2 & 0.43\%      & 0.39\%                & 36           \\
Choice 3 & 0.47\%    & 0.37\%                & 56           \\
Choice 4 & \textbf{0.80\%} & \textbf{0.81\%}        & 54           \\
Choice 5 & 0.76\%        & 0.48\%              & 85           \\
Choice 6 & 0.75\%        & 0.59\%               & \textbf{146} \\ \bottomrule
\end{tabular}
\end{table}

\cref{table:bucket_selection} shows the gains  of the {\algname-Binom} model and {\algname-Geo} model over the baseline model under these six definitions of bucket endpoints.
We also tabulate the number of buckets under these six definitions.
Since bucket endpoints need to be distinct, we remove duplicate values that repeat for different percentiles, and hence the number of buckets can be fewer than the 
arithmetically calculated counts.
For example, in Choice 1, using the 5-percentiles to define the bucket endpoints should ideally give 20 buckets, but removing the duplicates eventually gives us only 18 buckets.

In \cref{table:bucket_selection}, we see that Choice 4, a non-uniform bucket selection strategy, performs better in terms of consistently stronger model performance and relatively fewer number of buckets.
The comparison between Choice 4 and Choice 3, which have similar number of buckets, reveals that non-uniform bucket selection can be more effective than a uniform grid.
Further increasing bucket count, as in Choice 5 and Choice 6, may not provide convincing benefits to model performance, verifying our parsimonious principle in deciding bucket count.

\subsection{Online Experiment} \label{sec:online_exp}
We conduct an online A/B test by implementing our method onto the recommendation system on an real-world short video platform with hundreds of millions of daily-active users.
Specifically, our experiment focuses on the performance of our method on long videos, which are more relevant to the event of video watch time.

\subsubsection{Baselines.}
As in \cref{sec:offline_baseline}, our baseline  production model is a video watch time prediction model trained by the classical binary cross-entropy loss.

\subsubsection{Implementation Details.}
In our online A/B test, the traffic is split to different models uniformly, and the gain over the baseline model is reported.
The proposed method is applied in a re-ranker model where multiple criterion are considered simultaneously, and predicted video watch time serves as one critical component.

\subsubsection{Results.}

\begin{table}[]
\caption{Gain of our model over the production model on long videos in an online A/B test over five days. 
Reported metrics are accumulated  watch time and accumulated watch count.
For both metrics, the higher the numeric, the better.
The gain is calculated as a direct subtraction by the baseline model's result.
}
\label{table:online_main}
\setlength\tabcolsep{15pt}
\vspace{-0.5em}
\begin{tabular}{@{}ccc@{}}
\toprule
        & Watch Time & Watch Count \\\midrule
Day 1   & 4.167\%    & 3.262\%     \\
Day 2   & 6.160\%    & 5.541\%     \\
Day 3   & 6.175\%    & 7.683\%     \\
Day 4   & 4.446\%    & 3.752\%     \\
Day 5   & 4.436\%    & 4.297\%     \\\midrule
Average & 5.071\%    & 4.898\%    \\
\bottomrule
\end{tabular}
\end{table}

\cref{table:online_main} shows the results of online A/B test in each day over five days of traffic, on the metrics accumulated video watch time and accumulated watch counts, both of which are positive metrics, \ie, the higher the better. 
It is clear that our model provides significant gain on these two metrics in each day of the testing window.
This result coincides with our industrial offline experiments (\cref{sec:offline_exp}) and public datasets' experiments (\cref{sec:public_data}) that our method's predicted video watch time is more aligned with the ground-truth watch time.
Hence, our model can more accurately distinguish videos that attract the users most, and thus the recommendations based on our model increase users' video watch time and the number of watched videos (video watch count).

%% file: tex/conclusion.tex
\section{Conclusion}\label{sec:conclusion}
Video watch time prediction is one of the central problems in (short) video recommendation, as it directly affects the recommendation quality and  user experience.
Unlike conventional binary prediction tasks, this problem is significantly challenged by the complex user-video interaction; while most of the existing literature focus on black-box mechanical enhancement of the classical regression or classification losses.
In this paper, we for the first time  take on a  user-centric perspective to model video watch time, from which we propose a white-box statistical framework that translates our  domain knowledge on users' video-watching behavior into principled statistical models.
To better model users' fluctuating interests in continuing watching the video over the video horizon, we employ the bucketization strategy, which additionally helps to stabilize training objectives.
Extensive experiments of various nature validate our perspective and derived models, showing that our predictions align well with the ground-truth video watch time.

\section{Limitation and Future Work}\label{sec:limitation}
As an earliest effort in modeling video watch time through user behavior analysis, our paper biases towards succinct statistical models derived from relatively strict assumptions, such as Assumption~\ref{assump:binom_model} in \cref{sec:bimon_model}, so as to train and deploy models efficiently.
Nevertheless, our competitive experiment results validate our unique direction of modeling video watch time from the user-centric perspective and open the door for driving new statistical watch time models by softening and sophisticating those assumptions. 

 With accurate video watch time prediction, our {\algname} models can be directly connected  with explicit duration bias mitigation. 
 For example, one may utilize {\algname} model’s predictions without being overly sensitive to large predicted values, such as making video recommendations based on $\log(\text{predicted watch time} + 1)$ rather than the raw predicted watch time; or use our {\algname} framework to estimate a baseline model to contrast for duration debiasing.

Notice that our {\algname-Binom} and {\algname-Geo} are derived by respective user behavior assumptions, and that industrial recommendation platform typically contains various types of users.
One may therefore  cluster users and apply {\algname-Binom} and {\algname-Geo} correspondingly to each user segment, and/or use (learnable) personalized bucketization for each user, potentially with the help of reinforcement learning \citep{jointmatching2022,sdmgan2022,wmbrl2022} or reinforcement learning from human/AI feedback \citep{fantasticrewards2022,yang2023preferencegrounded,yang2024a}.

%% file: tex/appendix.tex
\appendix
\begin{center}
\LARGE	
\textbf{Appendix}
\end{center}

\section{The Expectation of {\algname-Geo} Model }\label{sec:proof_exp_bucket_geo}


Recall from \cref{sec:bucket_geo_model} that for a given set of continuation probabilities $\cbr{p_i}_{i=1}^{N+1}$, if the observation $t \in B_n = (x_{n-1}, x_n]$, the probability of $T=t$ is
\begin{equation}\label{eq:bucket_geometric_prob_appendix}\textstyle
    \Pr\br{T=t} = p_n^{t-x_{n-1}} \times (1-p_n)\times \prod_{i=1}^{n-1} p_i^{\Delta_i}\,,
\end{equation}
which also holds for the unbounded last bucket $(x_N, \infty)$.
By definition and \cref{eq:bucket_geometric_prob_appendix}, the expectation of $T$ is,
\begin{equation}\label{eq:bucket_geo_exp_derive}
    \begin{aligned}
        \E[T] &= \sum_{t = 0}^\infty t \times \Pr(T=t) \\
        &= \sum_{i=1}^{N+1} \underbrace{ \sum_{t=x_{i-1}+1}^{x_i} t \times \Pr(T=t)}_{=: \,\mu_i} \\
        &= \sum_{i=1}^{N+1} \mu_i
        \,.
    \end{aligned}
\end{equation}
For $\mu_i$, we have
\begin{equation}\label{eq:bucket_geo_exp_mu_i_derive}
    \begin{aligned}
        \mu_i &= \sum_{t=x_{i-1}+1}^{x_i} t \times \Pr(T=t) \\
        &= \sum_{t=x_{i-1}+1}^{x_i} t \times \prod_{j=1}^{i-1} p_j^{\Delta_j} \times {p_i^{t-x_{i-1}}(1-p_i)} \\
        &= \prod_{j=1}^{i-1} p_j^{\Delta_j}\, (1-p_i) \underbrace{\sum_{t=x_{i-1}+1}^{x_i} t\times p_i^{t-x_{i-1}}}_{=: \,\widetilde \mu_i} \\
        &= \prod_{j=1}^{i-1} p_j^{\Delta_j}\, (1-p_i)\,  \widetilde \mu_i
        \;.
    \end{aligned}
\end{equation}
For $\widetilde \mu_i$, we have
\begin{equation}\label{eq:bucket_geo_exp_tilde_mu_i_derive}
    \begin{aligned}
        \widetilde \mu_i &= \sum_{t=x_{i-1}+1}^{x_i} t\times p_i^{t-x_{i-1}} \\
        &= (x_{i-1} + 1)p_i^1 + (x_{i-1} + 2)p_i^2 + \cdots + x_i p_i^{\Delta_i} \\
        p_i \widetilde \mu_i &= (x_{i-1} + 1)p_i^2 + \cdots + (x_i - 1)p_i^{\Delta_i} + x_i p_i^{\Delta_i + 1} \\
        \implies (1-p_i)  \widetilde \mu_i &= x_{i-1}p_i + p_i^1 + p_i^2 + \cdots + p_i^{\Delta_i} - x_i p_i^{\Delta_i+1} \\
        &= x_{i-1}p_i + \frac{p_i\br{1-p_i^{\Delta_i}}}{1-p_i} - x_i p_i^{\Delta_i+1}
        \;.
    \end{aligned}
\end{equation}
Putting \cref{eq:bucket_geo_exp_tilde_mu_i_derive} back to \cref{eq:bucket_geo_exp_mu_i_derive} and with the notation that $p_0 = 1$, we have
\begin{equation}\label{eq:bucket_geo_exp_mu_i_derive_simplify}
    \forall\, i\,,\; \mu_i = \prod_{j=1}^{i-1} p_j^{\Delta_j}\, \br{x_{i-1}p_i + \frac{p_i\br{1-p_i^{\Delta_i}}}{1-p_i} - x_i p_i^{\Delta_i+1}}\,.
\end{equation}
With notational assumptions that $p_{N+1}^\infty = 0, x_{N+1} p_{N+1}^{\infty} = \infty  p_{N+1}^{\infty} = 0$, \cref{eq:bucket_geo_exp_mu_i_derive_simplify} also holds for the unbounded last bucket $(x_N, \infty)$, since
\begin{equation*}
\resizebox{0.47\textwidth}{!}{%
    $
    \begin{aligned}
        \widetilde \mu_{N+1} &= \sum_{t=x_{N}+1}^{\infty} t\times p_{N+1}^{t-x_{N}} \\
        &= (x_N + 1) p_{N+1}^1 + (x_N + 2)p_{N+1}^2 + \cdots \\
        \implies (1-p_{N+1})\widetilde \mu_{N+1} &= x_N p_{N+1}+p_{N+1}^1 + p_{N+1}^2 + \cdots \\
        &= x_N p_{N+1} + \frac{p_{N+1}}{ 1 - p_{N+1}} \\
        &= x_N p_{N+1} + \frac{p_{N+1}\br{1-p_{N+1}^\infty}}{ 1 - p_{N+1}} - x_{N+1}p_{N+1}^\infty
        \;.
    \end{aligned}
$%
}    
\end{equation*}
Finally, putting \cref{eq:bucket_geo_exp_mu_i_derive_simplify} back to \cref{eq:bucket_geo_exp_derive}, the expectation of $T$ under the bucketized geometric model/distribution ({\algname-Geo}) is
\begin{equation*}
    \E[T] = \sum_{i=1}^{N+1} \prod_{j=1}^{i-1} p_j^{\Delta_j}\, \br{x_{i-1}p_i + \frac{p_i\br{1-p_i^{\Delta_i}}}{1-p_i} - x_i p_i^{\Delta_i+1}}\,.
\end{equation*}
In \cref{sec:bucket_geo_model}, we replace $\cbr{p_i}$ by the corresponding fitted values $\cbr{\widehat p_i}$, but carry on the same implicit notational assumptions, resulting in the $\widehat \mu_{\text{{\algname-Geo}}}[T]$ estimator in \cref{eq:bucket_geo_expect}.

\section{Details of the Baseline Methods in Public Dataset Experiments}\label{sec:details_public_dataset}

In \cref{sec:public_data}, we compare our method with WLR, D2Q, OR, and TPM. 
In this section, we briefly introduce these baseline methods.

\begin{itemize}[leftmargin=*]
    \item \textit{WLR}: Weighted logistic regression combines the prediction of video watch time with the classical logistic regression for click-through rate. In particular, the impressed and clicked videos are treated as positive samples and the corresponding watch time is used as the sample weight in the classification loss. The impressed but unclicked videos are treated as negative samples, whose sample weight is designated as $1$.
    As discussed in \cref{remark:wlr_inference_time}, in inference time, the expected watch time is \textit{approximated} by exponentiated logits, under the extra assumption of \textit{small click probabilities}, which is inappropriate to short video platforms since all videos are (re-)played automatically without the user's click action.
    \item \textit{D2Q}: Duration-Deconfounded Quantile regression model first splits the training samples into a prescribed number of groups based on video length.
    Then, a quantile regression model is fitted to each group individually, modeling the watch time quantiles within each group.
    In inference time, an estimate of the expected video watch time can be obtained by using the inverse quantile function -- the empirical cumulative distribution function of the watch time of training samples in the corresponding video length group.
    \item \textit{OR}: Ordinal Regression model transforms the range of video watch time into multiple ranks, each of which is assigned with a classifier predicting whether the true value is greater than that rank.
    \item \textit{TPM}: Tree-based Progressive regression Model models video watch time as a series of dependent  conditional classification tasks that are arranged into a tree structure. 
    In inference time, the expected video watch time can be obtained by first traversing the tree to find the corresponding leaf node, and then estimating the expectation by a simple model, such as the mean of training samples in that leaf node.
    Notably, the training objective of TPM explicitly includes the variance of the predicted video watch time, as a way to minimize the prediction uncertainty.
\end{itemize}